\documentclass
[showkeys,superscriptaddress,10pt,balancelastpage,nofootinbib,twocolumn,nobibnotes,fleqn]{revtex4}%
\usepackage[verbose,colorlinks,hyperindex,breaklinks=true,pdfusetitle,citecolor=blue]{hyperref}
\usepackage{amsfonts}
\usepackage{amsmath}
\usepackage{amssymb}
\usepackage{graphicx}
\usepackage{color}
\usepackage{epsfig}
\usepackage{dcolumn}
\begin{document}
\title{Gap opening in the zeroth Landau level in gapped graphene: Pseudo-Zeeman
splitting in an angular magnetic field}
\author{M. Tahir\footnote{m.tahir@uos.edu.pk}}
\affiliation{Department of Physics, University of Sargodha, Sargodha $40100$, Pakistan.}
\author{K. Sabeeh\footnote{ksabeeh@qau.edu.pk}}
\affiliation{Department of Physics, Quaid-i-Azam University, 45320, Islamabad, Pakistan.}
\begin{abstract}
We present a theoretical study of gap opening in the zeroth Landau level in
gapped graphene as a result of pseudo-Zeeman interaction. The applied magnetic
field couples with the valley pseudospin degree of freedom of the charge
carriers leading to the pseudo-Zeeman interaction. To investigate its role in
transport at the Charge Neutrality Point (CNP), we study the integer quantum
Hall effect (QHE) in gapped graphene in an angular magnetic field in the
presence of pseudo-Zeeman interaction. Analytical expressions are derived for
the Hall conductivity using Kubo-Greenwood formula. We also determine the
longitudinal conductivity for elastic impurity scattering in the first Born
approximation. We show that pseudo-Zeeman splitting leads to a minimum in the
collisional conductivity at high magnetic fields and a zero plateau in the
Hall conductivity. Evidence for activated transport at CNP is found from the
temperature dependence of the collisional conductivity.
\end{abstract}
\maketitle
\section{INTRODUCTION}
In recent years, the experimental realization of a stable single layer of
carbon atoms\cite{1,2} has stimulated much interest in the studies of its
unusual properties\cite{3,4}. This material known as graphene is a tightly
packed honeycomb lattice of carbon atoms. Graphene monolayer is a gapless
semiconductor with conical touching of electron and hole bands. The charge
carriers in this system obey a linear dispersion relation near the Dirac
point, which endows it with unique electronic properties. This difference in
the nature of the quasiparticles in graphene from conventional two-dimensional
(2D) electronic systems has given rise to a host of new and unusual phenomena.
Besides the fundamental interest in understanding the electronic properties of
graphene there are also serious efforts to build nanoelectronic devices from
graphene\cite{5,6,7}.

The quantum Hall measurements in graphene were one of the key tools providing
evidence that the quasiparticles in graphene are chiral, massless fermions
known as Dirac fermions\cite{8,9,10,11,12,13}. A key difference in the
integral quantum Hall effect in graphene compared to the standard effect in
conventional two-dimensional electron gas (2DEG) systems is the occurrence of
a zeroth Landau level (LL) state. There have been several experimental as well
as theoretical studies of transport at the CNP which required analyzing the
role of the zeroth LL state\cite{14,15,16,17,18,19}. The nature of the
splitting of the electronic states at the CNP at present remains unclear,
whether this splitting is due to Zeeman interactions, electron-electron or
electron-phonon interactions, asymmetric gap, gapless edge states or due to
valley splitting is still an open question. The results obtained in \cite{16}
were explained on the basis of field dependent splitting of the zeroth Landau
level. Breakdown of the QHE in graphene leading to two insulating regimes were
analyzed in \cite{17,18}. On the theoretical side, \cite{15} addressed the
role of the zeroth LL in the QHE in graphene using Laughlin's gauge argument.
The role of disorder in addition to the Landau level structure was discussed
in detail in \cite{19}. In \cite{14}, the splitting of the zeroth Landau level
was observed and it was attributed to lifting of the sublattice and spin symmetry.

We consider gapped graphene in the presence of an external magnetic field. Due
to the breaking of sublattice symmetry a gap opens in the energy spectrum at
the Dirac points. One of the mechanisms that can lead to sublattice symmetry
breaking is through a local asymmetric chemical or electrical environment
provided by a substrate, such as epitaxial graphene on SiC or BN substrate.
Band gaps of various magnitudes can be induced depending on the
substrate\cite{20,21}. This symmetry breaking can have important consequences
for transport at CNP as it can lead to pseudo-Zeeman splitting of the zeroth
level when contributions of both the valleys are considered. It has been shown
in \cite{23,30,31,32,33,40,43,45} that if Berry phase effects are taken into
account, for crystals with broken spatial inversion symmetry, the electrons
acquire an orbital magnetic moment as a result of the self rotation of the
Bloch electron wave packet. This applies to Dirac fermions in graphene with
staggered sublattice potential which breaks inversion symmetry where the
orbital magnetic moment is associated with the valley index and can lead to
the valley QHE. In addition, various mechanism for valley filtering and valley
polarization have been discussed in relation to electronic
devices\cite{33,40,41,42,43}.

Our focus is on electron transport at CNP in gapped graphene in a tilted
magnetic field when the sublattice symmetry of graphene is broken resulting in
an energy gap at the Dirac point. This requires that we consider valley QHE
which has been discussed earlier in \cite{22,33,40} and more recently in
\cite{23}. In \cite{22}, the authors relate the heights of the plateaux in the
Hall conductivity and the peaks in the diagonal conductivity to the size of
the bandgap and the\ amount of disorder in the system. A semiclassical
presentation of the valley QHE in graphene is given in \cite{23}. Here, we
present a full quantum mechanical transport theory for valley QHE and analyze
the effects of the external tilted magnetic field on the transport at CNP at
finite temperature in the presence of screened charged impurities. From the
very beginning of our calculation, we explicitly introduce the Zeeman coupling
of the external magnetic field with the valley pseudospin of the Dirac
fermions and diagonalize the graphene Hamiltonian in its presence. Moreover,
we consider an external angular magnetic field\cite{14,15} applied to the
system in order to highlight the role of the out-of-plane to in-plane magnetic
field in the splitting of the zeroth LL. Furthermore, the analysis of
magnetotransport in this work is performed in the presence of elastic
scattering due to charged impurities which are known to be the dominant
scattering mechanism\cite{24,25,26,27,28,29} in graphene on a substrate. In
addition, this is a finite temperature study where the role of temperature in
magnetotransport at CNP can be investigated. The electrical transport
coefficients have been obtained from the standard Kubo formula in the
self-consistent Born approximation\cite{8,9,30}.

In section II, we present the formulation of the problem and numerical
discussion of density of states for different magnetic field strengths with
varying tilt angle from out-of-plane to in-plane. Section III contains the
derivation of the Hall conductivity as a function of the tilted magnetic field
including discussion of numerical results where as in section IV we evaluate
the longitudinal conductivity as a function of tilt angle with discussion of
results. In section V, summary of the work is given followed by an appendix in
section VI.
\section{Formulation}
We consider Dirac fermions in graphene which is in the $x-y$-plane in the
presence of a tilted magnetic field and pseudo-Zeeman interaction. The
magnetic field \{($B_{x}$, $0$, $B_{z}$) = ($B\sin\theta$, 0, $B\cos\theta$)\}
is applied at an angle $\theta$ with the $z$-direction which is perpendicular
to the graphene plane. The effective Hamiltonian for Dirac fermions in gapped
graphene in a magnetic field\cite{31,32,33} can be expressed as (the speed of
light $c=1$ in the minimal substitution that follows)%
\begin{equation}
H^{\tau_{z}}=V_{F}[\sigma_{x}(p_{x}+eA_{x})\tau_{z}+\sigma_{y}(p_{y}%
+eA_{y})]+\Delta_{z}\sigma_{z}. \label{1}%
\end{equation}
Here $\tau_{z}=\pm1$ for valleys $K$ and $K^{\prime}$, $\left(  A_{x}%
,A_{y}\right)  $ are the components of the vector potential, $V_{F}$
characterizes the Fermi velocity of Dirac fermions. We identify $\Delta
_{z}=-\mu_{B}^{\ast}B_{z}$ as the pseudo-Zeeman term with $B_{z}=B\cos\theta$,
$\sigma=\{\sigma_{x},\sigma_{y},\sigma_{z}\}$ are the Pauli matrices, the
effective Bohr magneton is $\mu_{B}^{\ast}=\frac{e\hslash}{2m_{e}^{\ast}}%
$\ with the effective mass $m_{e}^{\ast}=2\Delta\hslash^{2}/3a^{2}t^{2}$. The
Bohr magneton and the effective mass are expressed in terms of the gap energy
$\Delta=0.28$ eV (for graphene on SiC), the nearest neighbour hopping energy
$t=2.82$ eV and the lattice constant $a=0.246$ nm with the result that the
effective Bohr magneton ($\mu_{B}^{\ast})$ can be 30 times larger than the
free electron Bohr magneton ($\mu_{B}$)\cite{23,33,40}. This also allows us to
ignore the real spin Zeeman term. The above Hamiltonian is the same as the one
obtained in\cite{45,46} in the absence of valley-orbit coupling.

The Hamiltonian $H^{\tau_{z}}$ for the two valleys ($K,K^{\prime}$) can be
written as
 \begin{equation}
 H^{\tau_{z}}=V_{F}\left(
 \begin{array}
 [c]{c}%
 \Delta_{z}/V_{F}\\
 p_{x}\tau_{z}+ip_{y}+ieA_{y}%
 \end{array}%
 \begin{array}
 [c]{c}%
 p_{x}\tau_{z}-ip_{y}-ieA_{y}\\
 -\Delta_{z}/V_{F}%
 \end{array}
 \right)  \label{2}%
 \end{equation}
where in the diagonal terms $\Delta_{z}=\pm\mu_{B}^{\ast}B_{z}$ represents the
potential asymmetry between A and B lattice sites, which opens an energy gap
at the CNP. We have employed the Landau gauge and expressed the vector
potential as $\overrightarrow{A}=(0,$ $B_{z}x-B_{x}z,$ $0)$. The last term in
the Hamiltonian given in Eq.(1) is regarded as the pseudospin Zeeman term
($\Delta_{z}=\mu_{B}^{\ast}B_{z}$)\cite{23,30,31,32,33,43}, where the valleys
$K$ and $K^{\prime}$ serve as pseudospin up ($+1$) and pseudospin ($-1$),
respectively. Equation (2) is expressed as%
\begin{widetext}
 \begin{equation}
H^{\tau_{z}} = V_{F}\left(\begin{array}{lr}\Delta_z/V_F & p_x\tau_z-i p_y- i e (B_z x - B_x z)\\
p_x\tau_z+i p_y + i e (B_z x - B_x z) & -\Delta_z/V_F
      \end{array}\right).
 \label{3}%
\end{equation}
\end{widetext}
To obtain the energy eigen solutions of the above equation, one can use the
eigenvalue equation for a given spinor%
\begin{equation}
\Psi(\mathbf{r})=\binom{\phi_{1}(\mathbf{r})}{\phi_{2}(\mathbf{r})}, \label{4}%
\end{equation}
as%
\begin{equation}
H\binom{\phi_{1}(\mathbf{r})}{\phi_{2}(\mathbf{r})}=E\binom{\phi
_{1}(\mathbf{r})}{\phi_{2}(\mathbf{r})}. \label{5}%
\end{equation}
It yields the following equations%
\begin{equation}
\Delta_{z}\phi_{1}(\mathbf{r})-iV_{F}\left(  ip_{x}\tau_{z}+p_{y}%
+e(B_{z}x-B_{x}z)\right)  \phi_{2}(\mathbf{r})=E\phi_{1}(\mathbf{r}) \label{6}%
\end{equation}%
\begin{equation}
iV_{F}\left(  -ip_{x}\tau_{z}+p_{y}+e(B_{z}x-B_{x}z)\right)  \phi
_{1}(\mathbf{r})-\Delta_{z}\phi_{2}(\mathbf{r})=E\phi_{2}(\mathbf{r}).
\label{7}%
\end{equation}
We try the ansatz%
\begin{equation}
\binom{\phi_{1}(\mathbf{r})}{\phi_{2}(\mathbf{r})}=\frac{1}{\sqrt{L_{y}}}%
\exp[ik_{y}y]\binom{\varphi_{1}(\mathbf{r})}{\varphi_{2}(\mathbf{r})},
\label{8}%
\end{equation}
where $L_{y}$ is the dimensions of the graphene monolayer in the
$y$-direction. From Eqs.(6,7 \& 8) we obtain the eigenvalues of the $n$th LL
as%
\begin{align}
E_{0}^{\tau_{z}}  &  =-\tau_{z}\Delta_{z}\text{, \ \ \ \ \ \ \ \ }%
n=0\text{\ \ }\label{9}\\
E_{s,n}^{\tau_{z}}  &  =s\sqrt{\hslash^{2}\omega_{D}^{2}2\left\vert
n\right\vert +(\tau_{z}\Delta_{z})^{2}}\text{ , \ \ \ \ \ \ \ }n\neq0\nonumber
\end{align}
with $s=\pm$ for electrons and holes. The corresponding eigenfunction is%
\begin{equation}
\Psi_{s,n,k_{y}}^{\tau_{z}=+1}(r)=\frac{e^{ik_{y}y}}{\sqrt{L_{y}}}\left(
\begin{array}
[c]{c}%
sc_{1}\varphi_{\left\vert n\right\vert -1}[\frac{(x+x_{0})}{l}]\\
c_{2}\varphi\left\vert _{n}\right\vert [\frac{(x+x_{0})}{l}]
\end{array}
\right)  . \label{10}%
\end{equation}
The $n=0$ Landau level (the zeroth LL) requires separate treatment. It lies
just at the top of the valence band and its amplitude is only at B sites with
energy $E_{0}^{K}=-\Delta_{z}$, $c_{1}=0$ and $c_{2}=1$. In the above equation
$-$ is for holes and $+$ for electrons\cite{31,32}, $x_{0}=l^{2}k_{y}$,
$l=\sqrt{\frac{\hslash}{eB\cos\theta}}$, $c_{1}=\sin(\alpha_{n}/2)$ with
$\sin(\alpha_{n})=-\frac{s\hslash\omega_{D}\sqrt{2\left\vert n\right\vert }%
}{\sqrt{\hslash^{2}\omega_{D}^{2}2\left\vert n\right\vert +(\tau_{z}\Delta
_{z})^{2}}}$, $c_{2}=\cos(\alpha_{n}/2)$ with $\cos(\alpha_{n})=\frac
{s\Delta_{z}}{\sqrt{\hslash^{2}\omega_{D}^{2}2\left\vert n\right\vert
+(\tau_{z}\Delta_{z})^{2}}}$. The coefficients $c_{1}$ and $c_{2}$ are
normalized such that $\left\vert c_{1}\right\vert ^{2}+\left\vert
c_{2}\right\vert ^{2}=1$. $\varphi_{n}[\frac{(x+x_{0})}{l}]=\sqrt{\frac
{1}{\sqrt{\pi}2^{n}n!l}}H_{n}(\frac{x+x_{0}}{l})\exp[-\frac{1}{2}%
(\frac{x+x_{0}}{l})^{2}]$, $H_{n}(x)$ are the Hermite polynomials, $\omega
_{D}=V_{F}\sqrt{\frac{eB\cos\theta}{\hslash}}$\ is the cyclotron frequency of
Dirac Fermions and $n$ is an integer.
Similarly, for the $K^{\prime}$ valley ($\tau_{z}=-1$), the Hamiltonian yields
the same eigenvalues as given in Eq. (9) with eigenfunctions%
\begin{equation}
\Psi_{s^{\prime},n,k_{y}}^{\tau_{z}=-1}(r)=\frac{e^{ik_{y}y}}{\sqrt{L_{y}}%
}\left(
\begin{array}
[c]{c}%
s^{\prime}c_{1}^{\prime}\varphi\left\vert _{n}\right\vert [\frac{(x+x_{0})}%
{l}]\\
c_{2}^{\prime}\varphi_{\left\vert n\right\vert -1}[\frac{(x+x_{0})}{l}]
\end{array}
\right)  . \label{11}%
\end{equation}
where $s^{\prime}=\pm$for electrons and holes, $c_{1}^{\prime}=\sin(\alpha
_{n}^{\prime}/2)$ with $\sin(\alpha_{n}^{\prime})=\frac{s\hslash\omega
_{D}\sqrt{2\left\vert n\right\vert }}{\sqrt{\hslash^{2}\omega_{D}%
^{2}2\left\vert n\right\vert +(\tau_{z}\Delta_{z})^{2}}}$, $c_{2}^{\prime
}=\cos(\alpha_{n}^{\prime}/2)$ with $\cos(\alpha_{n}^{\prime})=-\frac
{s\Delta_{z}}{\sqrt{\hslash^{2}\omega_{D}^{2}2\left\vert n\right\vert
+(\tau_{z}\Delta_{z})^{2}}}$ . The energy of the $n=0$ LL for the $K^{\prime}$
valley is $E_{0}^{\tau_{z}=-1}=\Delta_{z}$. In this case, the Landau level
$n=0$ lies just at the bottom of the conduction band and its amplitude is only
at the $A$ sites.
The Density Of States (DOS) is defined as%
\begin{equation}
D(\varepsilon)=\frac{A}{\pi l^{2}}\underset{n,s,\tau_{z}}{%
{\displaystyle\sum}
}\delta\left(  \varepsilon-E_{s,n}^{\tau_{z}}\right)  \text{ \ \ }\ n\neq0,
\label{12}%
\end{equation}
and for $n=0$,\ the above equation at CNP, is written as%
\[
D_{CNP}(\varepsilon)=\frac{A}{\pi l^{2}}\underset{\tau_{z}}{%
{\displaystyle\sum}
}\delta\left(  \varepsilon-E_{0}^{\tau_{z}}\right)  ,
\]
where $E_{0}^{\tau_{z}}=\mp\Delta_{z}$ for $K$ and $K^{\prime}$ valleys
respectively, and $A$ is the area of the sample. Assuming a Gaussian
broadening of width $\Gamma$, the DOS at CNP is expressed as%
\begin{equation}
D_{CNP}(\varepsilon)=\frac{2}{2\pi l^{2}}\underset{\tau_{z}}{%
{\displaystyle\sum}
}\frac{1}{\Gamma\sqrt{2\pi}}\exp\left[  -\frac{(\varepsilon-E_{0}^{\tau_{z}%
})^{2}}{2\Gamma^{2}}\right]  , \label{13}%
\end{equation}
where $\Gamma$\ is the Gaussian distribution broadening width of zero shift.
Similarly, one can evaluate the DOS for the other valley ($K^{\prime}$).
\begin{figure}[ht]
\begin{center}
\includegraphics[width=0.4\textwidth]{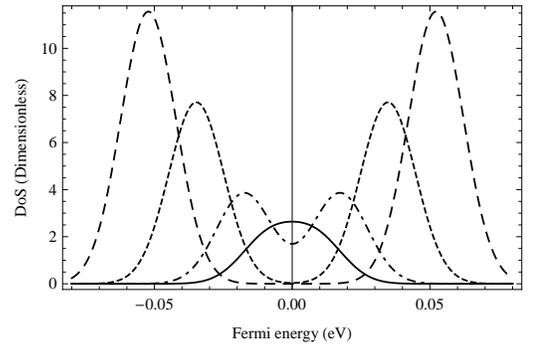} \caption{\label{dos} 
Density of States (DoS), dimensionless as a function of the Fermi
energy for different values of the magnetic field strength. The magnetic field
strength varied from 5 Tesla (solid line), 10 Tesla (dot-dashed line), 20
Tesla (dotted line) to 30 Tesla (dashed line) for fixed values of angle (zero
degree) and temperature (T = 0). } 
\end{center}
\end{figure}

The above expression for the density of states at CNP is plotted in Fig. (1)
as a function of the the Fermi energy (gate voltage) as the magnetic field
strength is varied to see the splitting of the zeroth Landau level. The
applied magnetic field is perpendicular to the graphene plane, the tilt angle
$\theta=0.$ The magnetic field is varied from 5 Tesla (solid line), 10 Tesla
(dot-dash line), 20 Tesla (dotted line), all the way to 30 Tesla (dashed
line). In Fig. (1), we see that as we increase the magnetic field strength,
the zeroth LL splits further apart. The gap in\ the density of states is well
resolved for high magnetic fields.
\begin{figure}[ht]
\begin{center}
\includegraphics[width=0.5\textwidth]{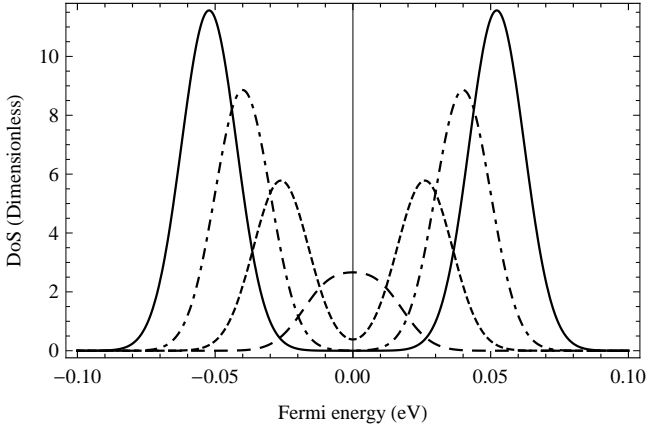} \caption{\label{dos2} 
Density of States (DoS), dimensionless as a function of the
Fermienergy for fixed value of the magnetic field strength at 30 Tesla. The
tiltangle is varied from zero (solid line), 40 (dot-dashed line), 60 (dotted
line) to 80 degree (dashed line) at zero temperature (T = 0). } 
\end{center}
\end{figure}

In Fig. (2), we plot the density of states
at CNP as a function of the Fermi energy for different values of the tilt
angle of the magnetic field (from our-of-plane to in-plane). The magnetic
field is fixed at 30 Tesla. The tilt angles are chosen to be $\theta=0$ (solid
line), $40$ (dot-dashed line), $60$ (dotted), and $80$ (dashed) degrees in
Fig. (2). The splitting of the zeroth Landau level washes out as we increase
the tilt angle of the magnetic field which is consistent with the
pseudo-Zeeman interpretation of this splitting. At $\theta=80$ degrees, when
the magnetic field is almost completely aligned with the graphene plane, the
gap has closed and a single peak with no splitting occurs at CNP. In Figs.(1
and 2) the following parameters were employed: $g=60$\cite{31,32}. The
broadening of the Landau levels generally depends on the magnetic field
strength, the temperature, Landau level index and the scattering parameters.
This requires a self consistent calculation usually performed numerically. In
order to carry out a tractable analytical calculation we have a chosen a
constant level width $\Gamma=10$ meV\cite{25,34,35,36}.

In the presence of a magnetic field, there are two contributions to
magnetoconductivity\cite{37,38}: the collisional (hopping) contribution and
the Hall contribution. The former is the localized state contribution which
carries the effects of SdH oscillations. The Hall contribution is the non
diagonal contribution. In order to calculate the electrical conductivity in
the presence of pseudo-Zeeman interactions and a tilted magnetic field we will
follow the formulation of\cite{39}, which is derived from the general
Liouville equation\cite{37,38} and includes dissipative effects. In the linear
response regime, the conductivity tensor is a sum of a diagonal and a
nondiagonal part : $\sigma_{\mu\nu}(\omega)=\sigma_{\mu\nu}^{d}(\omega
)+\sigma_{\mu\nu}^{nd}(\omega)$, $\mu,\nu=x,y$. In general, the diagonal
conductivity $\sigma_{\mu\nu}^{d}(\omega)=\sigma_{\mu\nu}^{diff}%
(\omega)+\sigma_{\mu\nu}^{\operatorname{col}}(\omega),$ accounts for both
diffusive and collisional contributions whereas the Hall contribution is
obtained from the nondiagonal conductivity $\sigma_{\mu\nu}^{nd}(\omega).$
Here, $\sigma_{xx}^{diff}=\sigma_{yy}^{diff}=0$ (here the diffusion
contribution is zero because the diagonal elements of the velocity operators
vanish) and $\sigma_{xx}^{\operatorname{col}}=\sigma_{yy}^{\operatorname{col}
}.$ This formulation has been employed successfully in electronic transport in
2DEG systems\cite{37,38} and more recently in graphene\cite{39}, and
references therein.
\section{Hall Conductivity}
The Hall conductivity $\sigma_{yx}$ is obtained from the nondiagonal elements
of the conductivity tensor, given by\cite{37,38,39}
\begin{equation}
\begin{split}
\sigma_{yx} &=\frac{2i\hbar e^{2}}{\Omega}\sum_{\xi\neq\xi^{\prime}}f_{\xi}(1-f_{\xi^{\prime}})
\left\langle \xi\right\vert v_{x}\left\vert
\xi^{\prime}\right\rangle \left\langle \xi^{\prime}\right\vert v_{y}\left\vert\xi\right\rangle \\
&\times \frac{(1-e^{\beta(E_{\xi}-E_{\xi^{\prime}})})}{\left(E_{\xi}-E_{\xi^{\prime}}\right)^{2}}.
\end{split}
 \label{14}
\end{equation}
Since $f_{\xi}(1-f_{\xi^{\prime}})(e^{\beta(E_{\xi}-E_{\xi^{\prime}})}%
)=f_{\xi^{\prime}}(1-f_{\xi})$ and $\Omega\rightarrow S_{0}\equiv L_{x}L_{y},$
we obtain%
\begin{equation}
\sigma_{yx}=\frac{2i\hslash e^{2}}{S_{0}}\underset{\xi\neq\xi^{\prime}}{\sum
}(f_{\xi}-f_{\xi^{\prime}})\frac{\left\langle \xi\right\vert v_{x}\left\vert
\xi^{\prime}\right\rangle \left\langle \xi^{\prime}\right\vert v_{y}\left\vert
\xi\right\rangle }{\left(  E_{\xi}-E_{\xi^{\prime}}\right)  ^{2}}. \label{15}%
\end{equation}
Since the $x$ and $y$ components of the velocity operator are $v_{x}%
=\frac{\partial H}{\partial p_{x}}$ and $v_{y}=\frac{\partial H}{\partial
p_{y}}$ where $H^{\tau_{z}}=V_{F}[\sigma_{x}p_{x}\tau_{z}+\sigma_{y}%
(p_{y}+eA_{y})]+\Delta_{z}\sigma_{z}.$ Therefore, $v_{x}=V_{F}\sigma_{x}%
\tau_{z}$ and $v_{y}=V_{F}\sigma_{y}$. Hence%
\begin{equation}
\left\langle \xi^{\prime}\right\vert v_{x}\left\vert \xi\right\rangle
=c_{1}c_{2}V_{F}(\delta_{n-1,n^{\prime}}+\delta_{n,n^{\prime}-1}) \label{16}%
\end{equation}
and%
\begin{equation}
\left\langle \xi\right\vert v_{y}\left\vert \xi^{\prime}\right\rangle
=-ic_{1}c_{2}V_{F}(\delta_{n^{\prime}-1,n}-\delta_{n-1,n^{\prime}}).
\label{17}%
\end{equation}
Since $\left\vert \xi\right\rangle \equiv\left\vert n,s,\tau_{z}%
,k_{y}\right\rangle $, there will be one summation over $k_{y}$\ which, with
periodic boundary conditions for $k_{y}$, will give%
\begin{equation}
\underset{k_{y}}{\sum}\rightarrow\frac{L_{y}}{2\pi}\overset{L_{x}/2l^{2}%
}{\underset{-L_{x}/2l^{2}}{\int}}dk_{y}=\frac{S_{0}}{2\pi l^{2}}. \label{18}%
\end{equation}
Substituting the values of the matrix elements of velocity in Eq. (15) yields%
\begin{equation}
\begin{split}
 \sigma_{yx} &=\frac{2\times2(c_{1}c_{2})^{2}\hbar e{{}^2}V_{F}{{}^2}}{2\pi l^{2}} \\
&\times \sum_{\xi\neq\xi^{\prime}}\frac{(f_{\xi}-f_{\xi
^{\prime}})\left[  \delta_{n,n^{\prime}-1}-\delta_{n-1,n^{\prime}}\right]
}{\left(  E_{\xi}-E_{\xi^{\prime}}\right) ^{2}}.
\end{split}
 \label{19}%
\end{equation}
Here factor of 2 is due to spin degeneracy. Since $E_{\xi}\equiv E_{s,n}%
^{\tau_{z}}=s\sqrt{\hslash^{2}\omega_{D}^{2}2\left\vert n\right\vert
+(\tau_{z}\Delta_{z})^{2}}$ we obtain%
 \begin{eqnarray}
\left( E_{\xi}-E_{\xi^{\prime}}\right)^{2} &=& [s\sqrt{2\hbar
 ^{2}\omega_{D}^{2}\lvert n\rvert +(\tau_{z}\Delta_{z})^{2}}\nonumber \\
&&- s^{\prime}\sqrt{2\hbar^{2}\omega_{D}^{2}\lvert n^{\prime}\rvert
+(\tau_{z}^{\prime}\Delta_{z})^{2}}]^{2}.
  \label{20}
 \end{eqnarray}
Substituting Eq. (20) in Eq. (19) we obtain the Hall conductivity%
\begin{widetext}
\begin{equation}
\sigma_{yx}=\frac{2(c_{1}c_{2})^{2}\hslash e{{}^2}V_{F}{{}^2}
}{2\pi l^{2}\hslash^{2}\omega_{D}^{2}}\underset{s,s^{\prime},n,n^{\prime}
,\tau_{z},\tau_{z}^{\prime}}{\sum}\frac{\left(  f_{s,n}^{\tau_{z}
}-f_{s^{\prime},n^{\prime}}^{\tau_{z}^{\prime}}\right)  \left[  \delta
_{n,n^{\prime}-1}-\delta_{n-1,n^{\prime}}\right]  }{\left[  s\sqrt{\left\vert
n\right\vert +(\tau_{z}\Delta_{z})^{2}/2\hslash^{2}\omega_{D}^{2}}-s^{\prime
}\sqrt{\left\vert n^{\prime}\right\vert +(\tau_{z}^{\prime}\Delta_{z}
)^{2}/2\hslash^{2}\omega_{D}^{2}}\right]  ^{2}} \label{21}
\end{equation}
\end{widetext}
The above equation can be further simplified and the final result for the
angular Hall conductivity (see Appendix for details) is%
\begin{equation}
\begin{split}
\sigma_{yx} &= \frac{2(c_{1}c_{2})^{2}e^{2}}{\pi\hbar}\sum_{n,j}
4 \left(n+\left(\frac{\Delta_{z}}{\hbar\omega_{D}\sqrt{2}}
\right)^{2} + \frac{1}{2} \right) \\
&\times \left(f_{+,n}^{j}-f_{+,n+1}^{j}+f_{-,n}^{j}-f_{-,n+1}^{j}\right)
\end{split}
\label{22}
\end{equation}
where we have introduced the sum over $j=\pm1$ for a concise final expression.
The effect of the tilted magnetic field can be seen in the distribution
function through $\omega_{D}=V_{F}\sqrt{\frac{eB\cos\theta}{\hslash}}$ as
$f_{n,s}^{j}=f(E_{n,s}^{j})=[\exp(\frac{s\sqrt{\hslash^{2}\omega_{D}%
^{2}2\left\vert n\right\vert +(j\Delta_{z})^{2}}-E_{F}}{k_{B}T}+1)]^{-1}$. The
Hall conductivity at CNP is written as
\begin{equation}
\begin{split}
\sigma_{yx} &= \frac{2(c_{1}c_{2})^{2}e^{2}}{\pi\hbar}\sum_{n=0,j}
4 \left(  \left(  \frac{\Delta_{z}}{\hslash\omega_{D}\sqrt{2}}\right)
^{2}+\frac{1}{2}\right) \\
&\times \left(f_{+,0}^{j}-f_{+,1}^{j}+f_{-,0}^{j}-f_{-,1}^{j}\right)
\end{split}
\nonumber
\end{equation}
where $f_{0}^{j}=f(E_{0}^{j})=[\exp(\frac{-j\Delta_{z}-E_{F}}{k_{B}T}%
+1)]^{-1}$. In the limit when the tilt angle ($\theta=0$) and the
Pseudo-Zeeman term vanishes ($\Delta_{z}=\mu_{B}^{\ast}B_{z}=0$), and we
consider transport contribution from a single valley only, the results
obtained are consistent with previous works in the
literature\cite{1,2,3,8,9,10,37}.
Elements of the resistivity tensor $\rho_{\mu\nu}$($\mu$,$\nu$=$x$,$y$) can be
determined from those of the conductivity tensor $\sigma_{\mu\nu}$, obtained
above, using the expressions: $\rho_{xx}=$ $\sigma_{yy}$ $/S$, $\rho_{yy}=$
$\sigma_{xx}$ $/S$ and $\rho_{xy}=$ $-\sigma_{yx}$ $/S$ where $S=$
$\sigma_{xx}$ $\sigma_{yy}-$ $\sigma_{xy}$ $\sigma_{yx}$ with $S\approx$
$\sigma_{xy}^{2}=n_{e}^{2}e^{2}/B^{2}\cos^{2}\theta$.
\begin{figure}[ht]
\begin{center}
\includegraphics[width=0.5\textwidth]{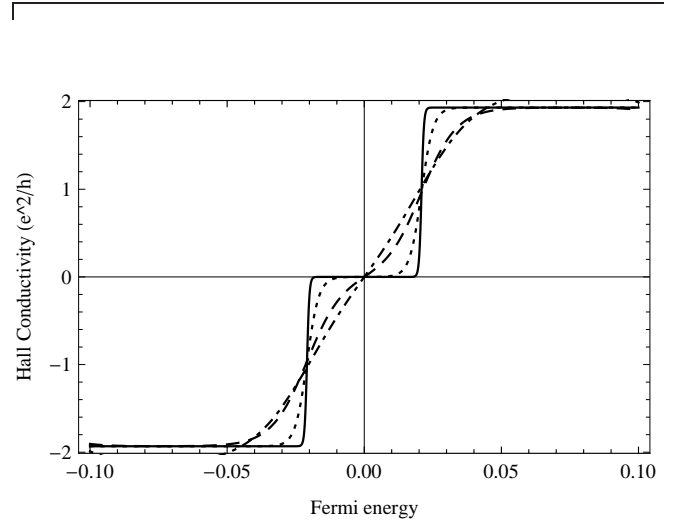} \caption{\label{dos3} 
Vanishing of the pleatues in the Hall conductivity as a function ofthe
Fermi energy with change in temperature. Temperature is varied from 5 (solid
line), 25 (dotted line), 75 (dashed line) to 125 K (dot-dashed line)for fixed
values of magnetic field (5 Tesla) and the Zeeman factor (g=60). } 
\end{center}
\end{figure}

The Hall conductivity, as a function of the Fermi energy, for different values
of the temperature is shown in Fig. (3). In this figure, we have shown results
for different values of temperature 5 (solid line), 25 (dotted line), 75
(dashed) and 125 K (dot-dashed). The step around CNP is washed out completely
at 125 K. The magnetic field is 5 Tesla, $g=60.$ This shows that in order to
observe pseudo-Zeeman splitting of the zeroth LL one needs to be in the regime
of high magnetic fields such that the thermal energy does not wash out the
splitting of the zeroth LL. Here, we have not considered the electron-phonon
interaction which can affect the results at high temperatures. That is
intended for future work. Furthermore, as the tilt angle of the magnetic field
is increased, the perpendicular component of the magnetic field becomes
smaller and the plateaus in the Hall conductivity disappear. We find that at a
temperature of 5K and magnetic field 5 Tesla, when $\theta=80$ degrees, the
steps disappear completely which is consistent with the discussion of the
density of states in section II.
\section{Collisional conductivity}
To obtain collisional contribution to conductivity, we assume that electrons
are elastically scattered by randomly distributed charged impurities as it has
been shown that charged impurities play a key role in the transport properties
of graphene near the Dirac point. This type of scattering is dominant at low
temperature. The collisional conductivity when spin degeneracy is considered
is given by \cite{37,38,39}%
\begin{equation}
\sigma_{xx}^{\operatorname{col}}=\frac{\beta e^{2}}{S_{0}}\underset{\xi
,\xi^{\prime}}{%
{\displaystyle\sum}
}f(E)(1-f(E^{\prime}))W_{\xi\xi^{\prime}}(E,E^{\prime})(x_{\xi}-x_{\xi
^{\prime}})^{2} \label{23}%
\end{equation}
where $f(E)=[\exp(\frac{E-E_{F}}{k_{B}T}+1)]^{-1}$ is the Fermi Dirac
distribution function with $f(E)=f(E^{\prime})$ for elastic scattering,
$k_{B}$ is the Boltzmann constant and $\mu$\ is the chemical potential.
$W_{\xi\xi^{\prime}}$ is the transmission rate between the one-electron states
$\left\vert \xi\right\rangle $ and $\left\vert \xi^{\prime}\right\rangle $,
$S_{0}$ the volume of the system, and $e$ the electron charge. Conduction
occurs by transitions through spatially separated states from $x_{\xi}$\ to
$x_{\xi^{\prime}}$, where $x_{\xi}=\left\langle \xi\right\vert x$ $\left\vert
\xi\right\rangle $ is the mean value of the $x$ component of the position
operator when the electron is in state $\left\vert \xi\right\rangle $. This is
the well known hopping type formula for transport in the presence of a
constant external magnetic field.
Collisional conductivity arises as a result of migration of the cyclotron
orbit due to scattering by charge impurities. The scattering rate $W_{\xi
\xi^{\prime}}$ is given by%
\begin{equation}
W_{\xi\xi^{\prime}}(E,E^{\prime})=\frac{2\pi N_{I}}{S_{0}\hslash}\underset{q}{%
{\displaystyle\sum}
}\left\vert U_{q}\right\vert ^{2}\left\vert F_{_{\xi\xi^{\prime}}%
}(u)\right\vert ^{2}\delta(E-E^{\prime})\delta_{k_{y},k_{y}^{\prime}+q_{y}}.
\label{24}%
\end{equation}
The Fourier transform of the screened impurity potential is%
\begin{equation}
U_{q}=U_{0}/\sqrt{q^{2}+k_{0}^{2}}, \label{25}%
\end{equation}
where $U_{0}=e^{2}/4\pi\epsilon_{0}\varepsilon$; $k_{0}$ is the screening wave
vector, $\varepsilon$ is the static dielectric constant of the material and
$\epsilon_{0}$ is the dielectric permittivity. $F_{_{\xi\xi^{\prime}}}(u)$ are
the form factors, $\left\langle \xi\right\vert e^{iq.r}\left\vert \xi^{\prime
}\right\rangle $\ with $u=l^{2}(q_{x}^{2}+q_{y}^{2})/2=\frac{q_{\perp}%
^{2}l^{2}}{2}$ with $q_{\perp}^{2}=(q_{x}^{2}+q_{y}^{2}).$ $N_{I}$ is the
impurity density and the wavefunction is $\left\vert \xi\right\rangle
\equiv\left\vert n,s,\tau_{z},k_{y}\right\rangle $. In the situation studied
here the diffusion contribution is zero because the diagonal elements of the
velocity operators vanish. Noting that $\sigma_{xx}^{\operatorname{col}%
}=\sigma_{yy}^{\operatorname{col}}$ and for screened impurity scattering such
that $k_{0}>>q$, we can ignore the $q$ dependence in Eq. (25).

Here $\left\langle \xi\right\vert x$ $\left\vert \xi\right\rangle =x_{0}$ is
the expectation value of the position with $(x_{\xi}-x_{\xi^{\prime}}%
)^{2}=(l^{2}q_{y})^{2}$ and $q_{y}=q_{\bot}\sin\zeta$. Since the wave function
oscillates around $-x_{0}$, we have%
\begin{equation}
\underset{k_{y}}{\sum}\rightarrow\frac{L_{y}}{2\pi}\overset{L_{x}/2l^{2}%
}{\underset{-L_{x}/2l^{2}}{\int}}dk_{y}=\frac{S_{0}}{2\pi l^{2}} \label{26}%
\end{equation}
and using cylindrical coordinates,%
\begin{equation}
\underset{q}{\text{ }\sum}\rightarrow\frac{S_{0}}{4\pi^{2}l^{2}}\overset{2\pi
}{\underset{0}{\int}}d\zeta\overset{\infty}{\underset{0}{\int}}du. \label{27}%
\end{equation}
The following matrix element between the two states can be evaluated to yield%
\begin{eqnarray}
 \left\vert \left\langle \xi\right\vert e^{iq.r}\left\vert \xi^{\prime
}\right\rangle \right\vert ^{2}  &=&\left\vert c_{2}^{2}F_{n,n}(u)+c_{1}
^{2}F_{n-1,n-1}(u)\right\vert ^{2}\label{28}\\
&=&e^{-u}\left[  c_{2}^{2}L_{n}\left(  u\right)  +c_{1}^{2}L_{n-1}\left(
u\right)  \right]  ^{2};n=n^{\prime},\nonumber
\end{eqnarray}
with%
\begin{equation}
\left\vert F_{n,n^{\prime}}(u)\right\vert ^{2}=\frac{n!}{n^{\prime}!}%
e^{-u}u^{n-n^{\prime}}\left[  L_{n^{\prime}}^{n-n^{\prime}}\left(  u\right)
\right]  ^{2};n^{\prime}\leq n, \label{29}%
\end{equation}
Inserting Eq. (24, 25, 26, 27 \& 28) in Eq.(23) the collisional conductivity
can be written as%
\begin{eqnarray}
\sigma_{xx}^{\operatorname{col}} &=&\frac{e^{2}}{h}\frac{2\beta N_{I}}
{l^{2}\hslash\omega_{D}}\underset{n,s,\tau_{z}}{\sum}\frac{U_{\circ}^{2}
}{k_{0}^{2}}\overset{\infty}{\underset{0}{\int}}due^{-u}u\left[  c_{2}
^{2}L_{n}\left(  u\right)  +c_{1}^{2}L_{n-1}\left(  u\right)  \right]
^{2} \nonumber \\
&&\times f(E_{n,s}^{\tau_{z}})(1-f(E_{n,s}^{\tau_{z}})) \label{30}
\end{eqnarray}
Finally, evaluating the above integral, we obtain the following result%
\begin{eqnarray}
\sigma_{xx}^{\operatorname{col}} &\approx& \frac{e^{2}}{h}\frac{2N_{I}U_{\circ
}^{2}\beta}{l^{2}k_{0}^{2}\hslash\omega_{D}}\underset{n,s,\tau_{z}}{\sum
}\left[  c_{2}^{4}(2n+1)+c_{1}^{4}(2n-1)+c_{1}^{2}c_{2}^{2}(2n)\right] \nonumber \\
&&\times f(E_{n,s}^{\tau_{z}})(1-f(E_{n,s}^{\tau_{z}})), \label{31}
\end{eqnarray}
where we have used the relation $\overset{\infty}{\underset{0}{\int}}%
due^{-u}u\left[  c_{2}^{2}L_{n}\left(  u\right)  +c_{1}^{2}L_{n-1}\left(
u\right)  \right]  ^{2}=c_{2}^{4}(2n+1)+c_{1}^{4}(2n-1)+c_{1}^{2}c_{2}%
^{2}(2n)$. In the limit of zero Zeeman interaction: $c_{2}^{4}$= $c_{1}%
^{4}=\frac{1}{4}$ and the integral will yield $6n$, which is consistent with
previous theoretical results\cite{8,9,10,39}. In the above expression, the
tilt angle of the magnetic field and the Zeeman interaction contribution
appears in the magnetic length $l=\sqrt{\frac{\hslash}{eB\cos\theta}}$ and the
distribution function $f(E_{n}^{\tau_{z}})=[\exp(\frac{s\sqrt{\hslash
^{2}\omega_{D}^{2}2\left\vert n\right\vert +(\tau_{z}\Delta_{z})^{2}}-E_{F}%
}{k_{B}T}+1)]^{-1}$ respectively.

At the CNP, where the contribution of the $n=0$ Landau level is crucial, the
collisional conductivity is expressed as%
\begin{equation}
\sigma_{xx}^{\operatorname{col}}\text{ (at CNP)}\varpropto\underset{\tau_{z}%
}{\sum}\beta f(E_{0}^{\tau_{z}})(1-f(E_{0}^{\tau_{z}})). \label{32}%
\end{equation}
\begin{figure}[ht]
\begin{center}
\includegraphics[width=0.5\textwidth]{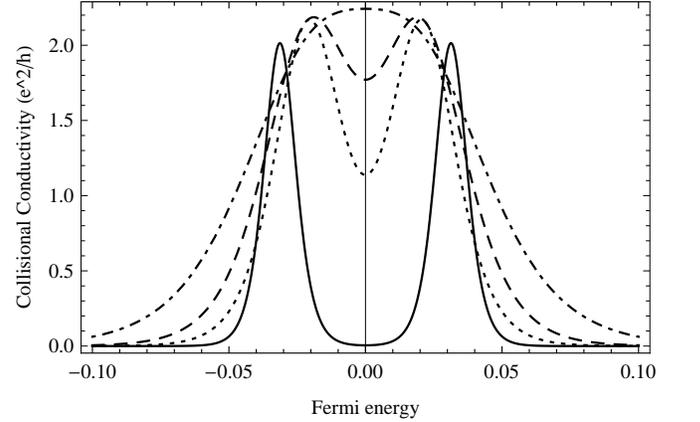} \caption{\label{dos4} 
Gap opening in the collisional conductivity as a function ofthe Fermi
energy at low temperature. Temperature is varied from5 K (solid line), 25 K
(dotted line), 75 K (dashed line), to 125 K(dot-dashed line) for fixed values
of the magnetic field (5 Tesla)and tilt angle (0 degree).} 
\end{center}
\end{figure}

The collisional conductivity\ at CNP given by Eq. (32) is shown graphically in
Fig (4) as a function of the Fermi energy for a fixed magnetic field at
$\theta=0$ tilt angle as the temperature (T) is varied: T= 5K (solid line),
25K (dotted line),75K (dashed line) and 125K (dot-dashed line) in Fig. (4).
The parameters\cite{16,17,18,19,24} used in all of our figures are :
$N_{I}=3\times10^{15}$ m$^{-2}$,\ $k_{0}=10^{-7}$m$^{-1}$, $V_{F}=10^{6}$ m/s,
magnetic field is 5 Tesla and $U_{0}$ = $e^{2}/4\pi\epsilon_{0}\varepsilon$.
We take $\varepsilon=4$ (graphene on a SiO$_{2}$ substrate) and $\epsilon_{0}$
the dielectric permittivity of free space$,$ with $k_{F}=(\pi n_{e})^{1/2}$
being the Fermi wave number. In Fig.(4), as the temperature is decreased, the
collisional conductivity exhibits a gap around CNP due to splitting of the
zeroth Landau level. This splitting at CNP is due to the pseudo- Zeeman
interaction as discussed earlier in the context of the density of states at
CNP. With an increase in temperature, at 195 K, the splitting is completely
washed out and there is only a single peak at CNP. Furthermore, Eq.(32) in the
limit of low temperatures or high magnetic fields, yields the temperature
dependence of the collisional conductivity as $\sigma_{xx}^{\operatorname{col}%
}$ (at CNP)$\varpropto\beta e^{-\beta\Delta_{z}}$ which represents an
activated type of behavior of the conductivity at the CNP.

In the limit of zero temperature ($T=0$), equation (32) can be expressed as%
\begin{equation}
\sigma_{xx}^{\operatorname{col}}\text{ (at CNP)}\varpropto\underset{\tau_{z}}{%
{\displaystyle\sum}
}\delta\left(  \varepsilon-E_{0}^{\tau_{z}}\right)  , \label{33}%
\end{equation}
which can then be written in terms of the Gaussian density of states as
derived and discussed in section II. The effect of the angular magnetic field
on the collisional conductivity follows the discussion presented in section II
for the density of states. Further, the results for the density of states at
CNP are consistent in the limit of no Zeeman interactions and a perpendicular
magnetic field (for $\theta=0)$ with the experimental as well as theoretical
results of \cite{34,35,36}.

Our results for both the Hall conductivity and the collisional conductivity
are relevant to transport measurements performed at the CNP on epitaxial
graphene grown on substrates such as SiC or BN where a band gap arises as a
result of interaction with the substrate. In this regard, we have shown that
one possible source of the opening of the gap in the density of states of the
zeroth Landau level is the pseudo-Zeeman interaction which leads to the
observed behavior of a plateau in the Hall conductivity and a dip in the
collisional conductivity at CNP. In gapless graphene, such as graphene on
SiO$_{2}$ substrate, opening of a gap in the density of states of the zeroth
Landau level can also occur due to valley splitting of the Landau levels. In
this case it can occur due to the inherent crystallographic symmetries of
graphene\cite{44}. The main difference in terms of realization in real
physical systems is that in gapped graphene the effective Bohr magneton or the
effective dipole moment that couples with the extenal magnetic field is much
larger which leads to larger valley splitting compared to gapless graphene for
the same magnetic field. For an energy gap of 0.28eV for graphene on SiC
substrate, the effective dipole moment is about 30 times larger than the free
electron spin magnetic moment. Therefore far smaller magnetic fields are
required to observe the valley splitting in gapped graphene as against gapless graphene.
\section{Summary}
In this work, we have investigated the coupling of an external magnetic field
with the valley pseudo spin of Dirac fermions and its effects on electron
transport in gapped graphene. Specifically, we have analyzed the splitting of
the zeroth LL due to this pseudo-Zeeman interaction and its effects on the
collisional and Hall conductivity at the CNP. To understand the role of the
pseudo-Zeeman interaction we have obtained analytic expressions and have
plotted the results for the density of states at CNP in the presence of an
external magnetic field whose tilt angle is varied. These results show that
the pseudo-Zeeman interaction causes splitting of the zeroth LL which vanishes
when the magnetic field is aligned along the graphene plane. We find that the
collisional conductivity at CNP shows activated behavior when the
pseudo-Zeeman interaction is taken into account. Furthermore, we are able to
show that as the temperature is increased for a fixed magnetic field, the
closing of the gap in the zeroth LL occurs.

\begin{widetext}
\section{Appendix}
 Equation (21) is written as%
\begin{align}
\sigma_{yx}  &  =\frac{2(c_{1}c_{2})^{2}\hslash e%
{{}^2}%
V_{F}%
{{}^2}%
}{2\pi l^{2}\hslash^{2}\omega_{D}^{2}}\underset{s,s^{\prime},n,n^{\prime}%
,\tau_{z},\tau_{z}^{\prime}}{\sum}\tag{A.1}\label{A.1}\\
&  \times\frac{\left(  f_{s,n}^{\tau_{z}}-f_{s^{\prime},n^{\prime}}^{\tau
_{z}^{\prime}}\right)  \left[  \delta_{n,n^{\prime}-1}-\delta_{n-1,n^{\prime}%
}\right]  }{\left[  s\sqrt{\left\vert n\right\vert +(\tau_{z}\Delta_{z}%
)^{2}/2\hslash^{2}\omega_{D}^{2}}-s^{\prime}\sqrt{\left\vert n^{\prime
}\right\vert +(\tau_{z}^{\prime}\Delta_{z})^{2}/2\hslash^{2}\omega_{D}^{2}%
}\right]  ^{2}}\nonumber
\end{align}
For $s,s^{\prime}$ = $+,+$ and $+,-,$ the summation in the above equation for
$n^{^{\prime}}=n+1$ is written as%
\begin{align}
&  =%
{\displaystyle\sum\limits_{n,\tau_{z},\tau_{z}^{\prime}}}
\frac{(f_{+,n}^{\tau_{z}}-f_{+,n+1}^{\tau_{z}^{\prime}})}{\left(
\sqrt{n+(\frac{\tau_{z}\Delta_{z}}{\hslash\omega_{D}\sqrt{2}})^{2}}%
-\sqrt{n+1+(\frac{\tau_{z}^{\prime}\Delta_{z}}{\hslash\omega_{D}\sqrt{2}}%
)^{2}}\right)  ^{2}}\tag{A.2}\label{A.2}\\
&  +\frac{(f_{+,n}^{\tau_{z}}-f_{-,n+1}^{\tau_{z}^{\prime}})}{\left(
\sqrt{n+(\frac{\tau_{z}\Delta_{z}}{\hslash\omega_{D}\sqrt{2}})^{2}}%
+\sqrt{n+1+(\frac{\tau_{z}^{\prime}\Delta_{z}}{\hslash\omega_{D}\sqrt{2}}%
)^{2}}\right)  ^{2}}.\nonumber
\end{align}
Note that $n^{^{\prime}}=n-1$ contribution vanishes as this corresponds to
transition to filled states. Equation (A.2) can be simplified to yield%
\begin{align}
&  =%
{\displaystyle\sum\limits_{n,\tau_{z},\tau_{z}^{\prime}}}
\left\{
\begin{array}
[c]{c}%
\left(  \sqrt{n+(\frac{\tau_{z}\Delta_{z}}{\hslash\omega_{D}\sqrt{2}})^{2}%
}+\sqrt{n+1+(\frac{\tau_{z}^{\prime}\Delta_{z}}{\hslash\omega_{D}\sqrt{2}%
})^{2}}\right)  ^{2}(f_{+,n}^{\tau_{z}}-f_{+,n+1}^{\tau_{z}^{\prime}})\\
+\left(  \sqrt{n+(\frac{\tau_{z}\Delta_{z}}{\hslash\omega_{D}\sqrt{2}})^{2}%
}-\sqrt{n+1+(\frac{\tau_{z}^{\prime}\Delta_{z}}{\hslash\omega_{D}\sqrt{2}%
})^{2}}\right)  ^{2}(f_{+,n}^{\tau_{z}}-f_{-,n+1}^{\tau_{z}^{\prime}})
\end{array}
\right\} \tag{A.3}\label{A.3}\\
&  /\left\{
\begin{array}
[c]{c}%
\left(  \sqrt{n+(\frac{\tau_{z}\Delta_{z}}{\hslash\omega_{D}\sqrt{2}})^{2}%
}-\sqrt{n+1+(\frac{\tau_{z}^{\prime}\Delta_{z}}{\hslash\omega_{D}\sqrt{2}%
})^{2}}\right)  ^{2}\\
\times\left(  \sqrt{n+(\frac{\tau_{z}\Delta_{z}}{\hslash\omega_{D}\sqrt{2}%
})^{2}}+\sqrt{n+1+(\frac{\tau_{z}^{\prime}\Delta_{z}}{\hslash\omega_{D}%
\sqrt{2}})^{2}}\right)  ^{2}%
\end{array}
\right\}  .\nonumber
\end{align}
For $s,s^{^{\prime}}=-,+$ and $s,s^{^{\prime}}=-,-$, the summation on the
right hand side of equation (A.1) is expressed as%
\begin{align}
&  =%
{\displaystyle\sum\limits_{n,\tau_{z},\tau_{z}^{\prime}}}
\frac{(f_{-,n}^{\tau_{z}}-f_{+,n+1}^{\tau_{z}^{\prime}})}{\left(
-\sqrt{n+(\frac{\tau_{z}\Delta_{z}}{\hslash\omega_{D}\sqrt{2}})^{2}}%
-\sqrt{n+1+(\frac{\tau_{z}^{\prime}\Delta_{z}}{\hslash\omega_{D}\sqrt{2}}%
)^{2}}\right)  ^{2}}\tag{A.4}\label{A.4}\\
&  +\frac{(f_{-,n}^{\tau_{z}}-f_{-,n+1}^{\tau_{z}^{\prime}})}{\left(
-\sqrt{n+(\frac{\tau_{z}\Delta_{z}}{\hslash\omega_{D}\sqrt{2}})^{2}}%
+\sqrt{n+1+(\frac{\tau_{z}^{\prime}\Delta_{z}}{\hslash\omega_{D}\sqrt{2}}%
)^{2}}\right)  ^{2}}\nonumber
\end{align}
Equation (A.4) can be simplified to yield%
\begin{align}
&  =%
{\displaystyle\sum\limits_{n,\tau_{z,}\tau_{z}^{\prime}}}
\left\{
\begin{array}
[c]{c}%
\left(  -\sqrt{n+(\frac{\tau_{z}\Delta_{z}}{\hslash\omega_{D}\sqrt{2}})^{2}%
}+\sqrt{n+1+(\frac{\tau_{z}^{\prime}\Delta_{z}}{\hslash\omega_{D}\sqrt{2}%
})^{2}}\right)  ^{2}(f_{-,n}^{\tau_{z}}-f_{+,n+1}^{\tau_{z}^{\prime}})\\
+\left(  -\sqrt{n+(\frac{\tau_{z}\Delta_{z}}{\hslash\omega_{D}\sqrt{2}})^{2}%
}-\sqrt{n+1+(\frac{\tau_{z}^{\prime}\Delta_{z}}{\hslash\omega_{D}\sqrt{2}%
})^{2}}\right)  ^{2}(f_{-,n}^{\tau_{z}}-f_{-,n+1}^{\tau_{z}^{\prime}})
\end{array}
\right\} \tag{A.5}\label{A.5}\\
&  /\left\{
\begin{array}
[c]{c}%
\left(  -\sqrt{n+(\frac{\tau_{z}\Delta_{z}}{\hslash\omega_{D}\sqrt{2}})^{2}%
}+\sqrt{n+1+(\frac{\tau_{z}^{\prime}\Delta_{z}}{\hslash\omega_{D}\sqrt{2}%
})^{2}}\right)  ^{2}\\
\times\left(  -\sqrt{n+(\frac{\tau_{z}\Delta_{z}}{\hslash\omega_{D}\sqrt{2}%
})^{2}}-\sqrt{n+1+(\frac{\tau_{z}^{\prime}\Delta_{z}}{\hslash\omega_{D}%
\sqrt{2}})^{2}}\right)  ^{2}%
\end{array}
\right\} \nonumber
\end{align}
Considering the numerator of Eq.(A.3), we obtain%
\begin{align}
&  =\left(  \sqrt{n+(\frac{\tau_{z}\Delta_{z}}{\hslash\omega_{D}\sqrt{2}}%
)^{2}}+\sqrt{n+1+(\frac{\tau_{z}^{\prime}\Delta_{z}}{\hslash\omega_{D}\sqrt
{2}})^{2}}\right)  ^{2}(f_{+,n}^{\tau_{z}}-f_{+,n+1}^{\tau_{z}^{\prime}%
})\tag{A.6}\label{A.6}\\
&  +\left(  \sqrt{n+(\frac{\tau_{z}\Delta_{z}}{\hslash\omega_{D}\sqrt{2}}%
)^{2}}-\sqrt{n+1+(\frac{\tau_{z}^{\prime}\Delta_{z}}{\hslash\omega_{D}\sqrt
{2}})^{2}}\right)  ^{2}(f_{+,n}^{\tau_{z}}-f_{-,n+1}^{\tau_{z}^{\prime}%
})\nonumber\\
&  =\left[  2n+(\frac{\tau_{z}\Delta_{z}}{\hslash\omega_{D}\sqrt{2}}%
)^{2}+(\frac{\tau_{z}^{\prime}\Delta_{z}}{\hslash\omega_{D}\sqrt{2}}%
)^{2}+1+2\sqrt{n+(\frac{\tau_{z}\Delta_{z}}{\hslash\omega_{D}\sqrt{2}})^{2}%
}\sqrt{n+1+(\frac{\tau_{z}^{\prime}\Delta_{z}}{\hslash\omega_{D}\sqrt{2}}%
)^{2}}\right] \nonumber\\
&  \times(f_{+,n}^{\tau_{z}}-f_{+,n+1}^{\tau_{z}^{\prime}})\nonumber\\
&  +\left[  2n+(\frac{\tau_{z}\Delta_{z}}{\hslash\omega_{D}\sqrt{2}}%
)^{2}+(\frac{\tau_{z}^{\prime}\Delta_{z}}{\hslash\omega_{D}\sqrt{2}}%
)^{2}+1-2\sqrt{n+(\frac{\tau_{z}\Delta_{z}}{\hslash\omega_{D}\sqrt{2}})^{2}%
}\sqrt{n+1+(\frac{\tau_{z}^{\prime}\Delta_{z}}{\hslash\omega_{D}\sqrt{2}}%
)^{2}}\right] \nonumber\\
&  \times(f_{+,n}^{\tau_{z}}-f_{-,n+1}^{\tau_{z}^{\prime}})\nonumber
\end{align}
and then simplifying the numerator of Eq. 5, we get%
\begin{align}
&  =\left(  -\sqrt{n+(\frac{\tau_{z}\Delta_{z}}{\hslash\omega_{D}\sqrt{2}%
})^{2}}+\sqrt{n+1+(\frac{\tau_{z}^{\prime}\Delta_{z}}{\hslash\omega_{D}%
\sqrt{2}})^{2}}\right)  ^{2}(f_{-,n}^{\tau_{z}}-f_{+,n+1}^{\tau_{z}^{\prime}%
})\tag{A.7}\label{A.7}\\
&  +\left(  -\sqrt{n+(\frac{\tau_{z}\Delta_{z}}{\hslash\omega_{D}\sqrt{2}%
})^{2}}-\sqrt{n+1+(\frac{\tau_{z}^{\prime}\Delta_{z}}{\hslash\omega_{D}%
\sqrt{2}})^{2}}\right)  ^{2}(f_{-,n}^{\tau_{z}}-f_{-,n+1}^{\tau_{z}^{\prime}%
})\nonumber\\
&  =\left[  2n+(\frac{\tau_{z}\Delta_{z}}{\hslash\omega_{D}\sqrt{2}}%
)^{2}+(\frac{\tau_{z}^{\prime}\Delta_{z}}{\hslash\omega_{D}\sqrt{2}}%
)^{2}+1-2\sqrt{n+(\frac{\tau_{z}\Delta_{z}}{\hslash\omega_{D}\sqrt{2}})^{2}%
}\sqrt{n+1+(\frac{\tau_{z}^{\prime}\Delta_{z}}{\hslash\omega_{D}\sqrt{2}}%
)^{2}}\right] \nonumber\\
&  \times(f_{-,n}^{\tau_{z}}-f_{+,n+1}^{\tau_{z}^{\prime}})\nonumber\\
&  +\left[  2n+(\frac{\tau_{z}\Delta_{z}}{\hslash\omega_{D}\sqrt{2}}%
)^{2}+(\frac{\tau_{z}^{\prime}\Delta_{z}}{\hslash\omega_{D}\sqrt{2}}%
)^{2}+1+2\sqrt{n+(\frac{\tau_{z}\Delta_{z}}{\hslash\omega_{D}\sqrt{2}})^{2}%
}\sqrt{n+1+(\frac{\tau_{z}^{\prime}\Delta_{z}}{\hslash\omega_{D}\sqrt{2}}%
)^{2}}\right] \nonumber\\
&  \times(f_{-,n}^{\tau_{z}}-f_{-,n+1}^{\tau_{z}^{\prime}})\nonumber
\end{align}
The denominator of Eq.(A.3) and Eq.(A.5) is%
\begin{equation}
\left(
\begin{array}
[c]{c}%
\left(  \sqrt{n+(\frac{\tau_{z}\Delta_{z}}{\hslash\omega_{D}\sqrt{2}})^{2}%
}-\sqrt{n+1+(\frac{\tau_{z}^{\prime}\Delta_{z}}{\hslash\omega_{D}\sqrt{2}%
})^{2}}\right)  ^{2}\\
\times\left(  \sqrt{n+(\frac{\tau_{z}\Delta_{z}}{\hslash\omega_{D}\sqrt{2}%
})^{2}}+\sqrt{n+1+(\frac{\tau_{z}^{\prime}\Delta_{z}}{\hslash\omega_{D}%
\sqrt{2}})^{2}}\right)  ^{2}%
\end{array}
\right)  =(\tau_{z}^{2}-\tau_{z}^{\prime2}-1)^{2}. \tag{A.8}\label{A.8}%
\end{equation}
One may notice that grouping terms such as $+,+$ and $+,-$ for $s$ and
$s^{\prime}$ that contain $f_{+,n}^{\tau_{z}}$ leads to the cancellation of
the following factor ($2\sqrt{n+(\frac{\tau_{z}\Delta_{z}}{\hslash\omega
_{D}\sqrt{2}})^{2}}\sqrt{n+1+(\frac{\tau_{z}^{\prime}\Delta_{z}}{\hslash
\omega_{D}\sqrt{2}})^{2}})$ in Eq. (A.6). The same holds for the $-,-$ and
$-,+$ terms in Eq. (A.7). Now using Eqs. (A.3), (A.5), (A.6), (A.7) and (A.8)
in Eq.(A.1), we arrive at the result
\begin{equation}
\sigma_{yx}=\frac{(c_{1}c_{2})^{2}e^{2}}{\pi\hbar}\times%
{\displaystyle\sum\limits_{n,\tau_{z},\tau_{z}^{\prime}}}
4\left[
\begin{array}
[c]{c}%
\left(  n+\frac{1}{2}(\frac{\tau_{z}\Delta_{z}}{\hslash\omega_{D}\sqrt{2}%
})^{2}+\frac{1}{2}(\frac{\tau_{z}^{\prime}\Delta_{z}}{\hslash\omega_{D}%
\sqrt{2}})^{2}+\frac{1}{2}\right) \\
(f_{+,n}^{\tau_{z}}-f_{+,n+1}^{\tau_{z}^{\prime}}+f_{-,n}^{\tau_{z}}%
-f_{-,n+1}^{\tau_{z}^{\prime}})
\end{array}
\right]  /(\tau_{z}^{2}-\tau_{z}^{\prime2}-1)^{2}. \tag{A.9}\label{A.9}%
\end{equation}
Performing the summation over $\tau_{z}$ and $\tau_{z}^{\prime}$ for $+,+$ and
$+,-$ respectively, Eq. (A.9) is simplified to yield%
\begin{equation}
\sigma_{yx}=\frac{(c_{1}c_{2})^{2}e^{2}}{\pi\hbar}\times\left\{
\begin{array}
[c]{c}%
{\displaystyle\sum\limits_{n}}
4\left[
\begin{array}
[c]{c}%
\left(  n+\frac{1}{2}(\frac{\Delta_{z}}{\hslash\omega_{D}\sqrt{2}})^{2}%
+\frac{1}{2}(\frac{\Delta_{z}}{\hslash\omega_{D}\sqrt{2}})^{2}+\frac{1}%
{2}\right) \\
(f_{+,n}^{+}-f_{+,n+1}^{+}+f_{-,n}^{+}-f_{-,n+1}^{+})
\end{array}
\right]  +\\%
{\displaystyle\sum\limits_{n}}
4\left[
\begin{array}
[c]{c}%
\left(  n+\frac{1}{2}(\frac{\Delta_{z}}{\hslash\omega_{D}\sqrt{2}})^{2}%
+\frac{1}{2}(\frac{\Delta_{z}}{\hslash\omega_{D}\sqrt{2}})^{2}+\frac{1}%
{2}\right) \\
(f_{+,n}^{+}-f_{+,n+1}^{-}+f_{-,n}^{+}-f_{-,n+1}^{-})
\end{array}
\right]
\end{array}
\right\}  . \tag{A.10}\label{A.10}%
\end{equation}
After simplifying the above equation we get%
\begin{equation}
\sigma_{yx}=\frac{(c_{1}c_{2})^{2}e^{2}}{\pi\hbar}\times\left\{
{\displaystyle\sum\limits_{n}}
4\left[
\begin{array}
[c]{c}%
\left(  n+(\frac{\Delta_{z}}{\hslash\omega_{D}\sqrt{2}})^{2}+\frac{1}%
{2}\right)  \times\\
\left(  (2f_{+,n}^{+}-f_{+,n+1}^{+}+2f_{-,n}^{+}-f_{-,n+1}^{+})-(f_{+,n+1}%
^{-}+f_{-,n+1}^{-})\right)
\end{array}
\right]  \right\}  . \tag{A.11}\label{A.11}%
\end{equation}
Similarly performing the summation over $\tau_{z}$ and $\tau_{z}^{\prime}$ for
$-,+$ and $-,-,$ Eq. (A.9) is simplified to yield%
\begin{equation}
\sigma_{yx}=\frac{(c_{1}c_{2})^{2}e^{2}}{\pi\hbar}\times%
{\displaystyle\sum\limits_{n}}
4\left[
\begin{array}
[c]{c}%
\left(  n+(\frac{\Delta_{z}}{\hslash\omega_{D}\sqrt{2}})^{2}+\frac{1}%
{2}\right)  \times\\
\left(  (2f_{+,n}^{-}-f_{+,n+1}^{+}+2f_{-,n}^{-}-f_{-,n+1}^{+})-(f_{+,n+1}%
^{-}+f_{-,n+1}^{-})\right)
\end{array}
\right]  . \tag{A.12}\label{A.12}%
\end{equation}
Finally combining Eqs. (A.11 and A.12), we arrive at the final result for the
Hall conductivity%
\begin{equation}
\sigma_{yx}=\frac{2(c_{1}c_{2})^{2}e^{2}}{\pi\hbar}\times\left\{
{\displaystyle\sum\limits_{n,j}}
4\left[
\begin{array}
[c]{c}%
\left(  n+(\frac{\Delta_{z}}{\hslash\omega_{D}\sqrt{2}})^{2}+\frac{1}%
{2}\right)  \times\\
(f_{+,n}^{j}-f_{+,n+1}^{j}+f_{-,n}^{j}-f_{-,n+1}^{j})
\end{array}
\right]  \right\}  \tag{A.13}\label{A.13}%
\end{equation}
with $j=\pm1.$ In the limit $\Delta_{z}=0,$ the above result reduces to that
of \cite{39} exactly.
\end{widetext}
\section{Acknowledgement}
K. Sabeeh would like to acknowledge the support of the Higher Education
Commission (HEC) of Pakistan through project No. 20-1484/R\&D/09 and the Abdus
Salam International Center for Theoretical Physics (ICTP) for support through
the Associateship Scheme.

\end{document}